\documentclass[a4paper, 12pt, oneside]{article}
\usepackage[cp1251]{inputenc}
\usepackage[english, russian]{babel}
\usepackage{ graphics,amsfonts, amsmath,bm,amssymb, array, eucal}
\usepackage[dvips]{graphicx}
\usepackage[flushleft,small]{caption2}

\makeatletter
\renewcommand{\@biblabel}[1]{#1.\hfill}

\newcommand{\const}{\mathop{\rm const\, }}
\renewcommand{\Re}{\mathop{\rm Re\,}}
\renewcommand{\Im}{\mathop{\rm Im\,}}
\makeatother
\renewcommand{\baselinestretch}{1.2}

\begin{document}
\newcommand{\mc}[1]{\mathcal{#1}}
\newcommand{\E}{\mc{E}}
\thispagestyle{empty}
\large

\renewcommand{\abstractname}{Abstract }
\renewcommand{\refname}{\begin{center} REFERENCES\end{center}}

 \begin{center}
\bf Longitudinal dielectric permeability into quantum degenerate
plasma with frequency of collisions proportional to the
module of a wave vector
\end{center}\medskip
\begin{center}
  \bf A. V. Latyshev\footnote{$avlatyshev@mail.ru$} and
  A. A. Yushkanov\footnote{$yushkanov@inbox.ru$}
\end{center}\medskip

\begin{center}
{\it Faculty of Physics and Mathematics,\\ Moscow State Regional
University, 105005,\\ Moscow, Radio str., 10--A}
\end{center}\medskip

\begin{abstract}
Formulas for the longitudinal dielectric permeability in quantum
degenerate collisional plasma with the frequency of
collisions proportional to the module of the wave
vector, in approach Мермина, are received.
Equation of Shr\"{o}dinger---Boltzmann with integral
of collisions relaxation type in Mermin's appro\-ach is applied.

It is spent numerical and graphic
comparison of the real and imaginary parts of dielectric function
of non-degenerate and maxwellian collisional quantum plasma
with a constant and a variable frequencies of collisions.
It is shown, that the longitu\-dinal
dielectric function weakly depends on a wave vector.


\end{abstract}

\begin{center}
{\bf 1. Introduction}
\end{center}

In  Klimontovich and Silin's work  \cite{Klim} expression
for longitudinal and trans\-verse dielectric permeability of quantum
collisionless plasmas has been re\-cei\-ved.

Then in Lindhard's work \cite{Lin} expressions
has been received  also for the same characteristics of quantum
collisionless plasma.

By Kliewer and Fuchs \cite{Kliewer} it has been shown, that
direct generalisation of formulas of Lindhard  on a case of collisionless
plasmas, is incorrectly.
This lack for the longitudinal dielectric
permeability has been eliminated in work of Mermin \cite{Mermin} for
collisional plasmas.
In this work of Mermin \cite{Mermin} on the basis of the analysis
of a nonequilibrium matrix
density in $ \tau $-approach expression for
longitudinal dielectric permeability of quantum collisional plasmas
in case of constant frequency of collisions of particles of plasma
has been announced.

For collisional plasmas correct formulas longitudinal and transverse
electric conductivity and dielectric permeability are received
accordingly in works \cite{Long} and \cite{Trans}. In these works
kinetic  Wigner---Vlasov---Boltzmann equation
in relaxation approximation in coordinate space was used.

In work \cite{Trans2} the formula for the transverse electric
conductivity of quantum collisional plasmas with use of the kinetic
Shr\"{o}dinger---Boltzmann equation in Mermin's approach  (in space of
momentum) has been deduced.

In work \cite{Long2} the formula for the longitudinal dielectric
permeability of quantum collisional plasmas with use of the kinetic
Shr\"{o}dinger---Boltzmann equation in approach of Mermin (in space of
momentum) with any variable frequency of collisions depending from
wave vector  has been deduced.

In the present work on the basis of results from our previous work
\cite{Long2} formulas for longitudinal dielectric permeability
in quantum degenerate colli\-sio\-nal plasma with frequency of collisions,
proportional to the module of a wave vector are received.
The modelling is thus used Shr\"{o}dinger---Boltzmann equation
in relaxation approximation.

In our work \cite{Lat2007} formulas for longitudinal and transverse
electric conductivity in the classical collisional
gaseous (maxwellian) plasma with frequency of collisions
of plasma particles proportional to the
module particles velocity  have been deduced.

Research of
skin-effect in classical collisional gas plasma with frequency
of collisions proportional to the module particles velocity
has been carried out in work \cite{Lat2006}.

Let's notice, that interest to research of the phenomena
in quantum plasma grows in last years \cite{Manf} -- \cite{Ropke}.

\begin{center}
\bf 1.  Longitudinal dielectric function of quantum collisional
plasma with variable collisional frequency
\end{center}

In work \cite{Long} longitudinal dielectric function of the quantum
collisional plasmas with frequency of collisions,
proportional to the module of a wave vector has been received
$$
\varepsilon_l({\bf q},\omega,\nu)=1+\dfrac{4\pi e^2}{q^2}\Big[B({\bf q},\omega+
i \bar \nu)+ \hspace{6cm}
$$
$$+ib_{\bar \nu}({\bf q},\omega+i \bar \nu)
\dfrac{b({\bf q},0)-b({\bf q},\omega+i\bar \nu)}
{\omega b({\bf q},0)+ib_{\omega,\bar \nu}({\bf q},\omega+i\bar \nu)}\Big].
\eqno{(1.1)}
$$\medskip

In the formula (1.1) $e $ is the electron charge, $ {\bf q} $ is the wave
vector, $ \omega $ is the frequency of oscillations of an electromagnetic field,
$ \nu ({\bf k}) $ is the frequency of collisions of particles of
plasma,
$$
\bar\nu=\bar \nu({\bf k,q})=\bar \nu({\bf k}+\dfrac{{\bf q}}{2},
{\bf k}-\dfrac{{\bf q}}{2})=\dfrac{\nu\big({\bf k}+\dfrac{{\bf q}}{2}\big)+
\nu\big({\bf k}-\dfrac{{\bf q}}{2}\big)}{2},
\eqno{(1.2)}
$$\medskip

$$
B({\bf q},\omega+i\bar\nu)=\int\dfrac{d^3k}{4\pi^3}\Big(f_{{\bf k+q}/2}-
f_{{\bf k-q}/2}\Big)\Xi(\omega+i\bar \nu({\bf k+q}/2,{\bf k-q}/2)),
\eqno{(1.3)}
$$\medskip
$$
b({\bf q},\omega+i\bar\nu)=\int\dfrac{d^3k}{4\pi^3}\Big(f_{{\bf k+q}/2}-
f_{{\bf k-q}/2}\Big)\Xi(\omega+i\bar \nu({\bf k+q}/2,{\bf k-q}/2))\times
$$\medskip
$$
\times\dfrac{\bar \nu({\bf k+q}/2,{\bf k-q}/2)}{\omega+i
\bar \nu({\bf k+q}/2,{\bf k-q}/2)},
\eqno{(1.4)}
$$\medskip
$$
b({\bf q},0)=\int\dfrac{d^3k}{4\pi^3}\Big(f_{{\bf k+q}/2}-
f_{{\bf k-q}/2}\Big)\Xi(0)\dfrac{\bar \nu({\bf k+q}/2,{\bf k-q}/2)}{\omega+i
\bar \nu({\bf k+q}/2,{\bf k-q}/2)},
\eqno{(1.5)}
$$\medskip
$$
b_{\bar \nu}({\bf q},\omega+i\bar\nu)=
\int\dfrac{d^3k}{4\pi^3}\Big(f_{{\bf k+q}/2}-
f_{{\bf k-q}/2}\Big)\Xi(\omega+i\bar \nu({\bf k+q}/2,{\bf k-q}/2))\times
$$\medskip
$$
\times{\bar \nu({\bf k+q}/2,{\bf k-q}/2)},
\eqno{(1.6)}
$$\medskip
$$
b_{\omega,\bar \nu}({\bf q},\omega+i\bar\nu)=
\int\dfrac{d^3k}{4\pi^3}\Big(f_{{\bf k+q}/2}-
f_{{\bf k-q}/2}\Big)\Xi(\omega+i\bar \nu({\bf k+q}/2,{\bf k-q}/2))\times
$$\medskip
$$
\times\dfrac{\bar \nu^2({\bf k+q}/2,{\bf k-q}/2)}{\omega+i
\bar \nu({\bf k+q}/2,{\bf k-q}/2)},
\eqno{(1.7)}
$$\medskip

In integrals (1.3) -- (1.7) the following designations are accepted
$$
\Xi(\omega+i\bar \nu({\bf k+q}/2,{\bf k-q}/2))=$$$$=
\dfrac{1}{\E_{{\bf k-q}/2}-\E_{{\bf k+q}/2}+\hbar[\omega+i \bar
 \nu({\bf k+q}/2,{\bf k-q}/2)]},
$$\medskip
$$
f_{{\bf k}}=\dfrac{1}{1+\exp\Big(\dfrac{\E_{{\bf k}}-\mu}{k_BT}\Big)},
$$\medskip
$$
\E_{{\bf k\pm q}/2}=\dfrac{\hbar^2}{2m}\Big({\bf k}\pm\dfrac{{\bf
q}}{2}\Big)^2.
$$

Here $m $ is the electron mass, $k_B $ is the Boltzmann constant,
$ \mu $ is the chemical potential of molecules of gas, $ \hbar $ ie the
Planck's constant.

Let's show, that at $ \nu ({\bf k}) = \nu =\const $, i.e. at a constant
collisional frequency the formula (1.1) passes in the known Mermin's formula
\cite{Mermin}

$$
\varepsilon_l^{\rm Mermin}=1+\dfrac{4\pi e^2}{q^2}\dfrac{(\omega+i \nu)
B({\bf q},\omega+i \nu)B({\bf q},0)}{\omega B({\bf q},0)+i \nu
B({\bf q},\omega+i \nu)}.
\eqno{(1.8)}
$$

In (1.8) the following designations are used

$$
B({\bf q},\omega+i \nu)=\int \dfrac{d^3k}{4\pi^3}
(f_{{\bf k+q}/2}-f_{{\bf k-q}/2})\Xi(\omega+ i \nu),
\eqno{(1.9)}
$$
$$
B({\bf q},0)=\int \dfrac{d^3k}{4\pi^3}
(f_{{\bf k+q}/2}-f_{{\bf k-q}/2})\Xi(0),\qquad
$$
$$
\Xi(\omega+i \nu)=\dfrac{f_{{\bf k+q}/2}-f_{{\bf k-q}/2}}
{\E_{{\bf k-q}/2}-\E_{{\bf k+q}/2}+\hbar(\omega+i \nu)}.
$$

Let's notice, that at $ \nu({\bf k}) \equiv \nu $, $ \bar\nu({\bf k, q}) \equiv
\nu $, and we receive following equalities
$$
B({\bf q},\omega+i\bar \nu)\equiv B({\bf q},\omega+i \nu),
$$
$$
b({\bf q},\omega+i\bar \nu)=\dfrac{\nu}{\omega+i \nu}B({\bf q},\omega+i \nu),
$$
$$
b({\bf q},0)=\dfrac{\nu}{\omega+i \nu}B({\bf q},0),
$$
$$
b_{\bar \nu}({\bf q},\omega+i\bar \nu)=\nu B({\bf q},\omega+i \nu),
$$
$$
b_{\omega,\bar \nu}({\bf q},\omega+i\bar \nu)=\dfrac{\nu^2}{\omega+i \nu}
B({\bf q},\omega+i \nu).
$$

It is as a result received, that
$$
\varepsilon_l({\bf q},\omega,\nu)=1+\dfrac{4\pi e^2}{q^2}
B({\bf q},\omega+i \nu)\Big[1+$$$$+i \nu\dfrac{B({\bf q},0)-
B({\bf q},\omega+i \nu)}{\omega B({\bf q},0)+
i \nu B({\bf q},\omega+i \nu)}\Big] \equiv \varepsilon_l^{\rm Mermin}
({\bf q},\omega,\nu).
$$

Each of integrals (1.3) -- (1.7) we will break into a difference
of two integrals. In each of two integrals it is realizable the obvious
linear replacement of variables. It is as a result received, that
$$
B({\bf q},\omega+i\bar \nu)=\int \dfrac{d^3k}{4\pi^3}f_{{\bf k}}
\Big[\Xi(\omega+i\bar\nu({\bf k,k-q}))-\Xi(\omega+i\bar\nu({\bf k+q,k}))\Big],
\eqno{(1.10)}
$$

$$
b({\bf q},\omega+i\bar \nu)=\int \dfrac{d^3k}{4\pi^3}f_{{\bf k}}
\Big[\dfrac{\bar\nu({\bf k,k-q})}{\omega+i\bar\nu({\bf k,k-q})}
\Xi(\omega+i\bar\nu({\bf k,k-q}))-$$$$-\dfrac{\bar\nu({\bf k+q,k})}
{\omega+i\bar\nu({\bf k+q,k}))}
\Xi(\omega+i\bar\nu({\bf k+q,k}))\Big],
\eqno{(1.11)}
$$

$$
b({\bf q},0)=\int \dfrac{d^3k}{4\pi^3}f_{{\bf k}}
\Big[\dfrac{\bar\nu({\bf k,k-q})}{\omega+i\bar\nu({\bf k,k-q})(\E_{{\bf k-q}}-
\E_{{\bf k}})}
-$$$$-\dfrac{\bar\nu({\bf k+q,k})}
{\omega+i\bar\nu({\bf k+q,k}))(\E_{\bf k}-\E_{{\bf k+q}})}\Big],
\eqno{(1.12)}
$$

$$
b_{\bar \nu}({\bf q},\omega+i\bar \nu)=\int \dfrac{d^3k}{4\pi^3}f_{{\bf k}}
\Big[{\bar\nu({\bf k,k-q})}
\Xi(\omega+i\bar\nu({\bf k,k-q}))-$$$$-{\bar\nu({\bf k+q,k})}
\Xi(\omega+i\bar\nu({\bf k+q,k}))\Big],
\eqno{(1.13)}
$$

$$
b_{\omega,\bar\nu}({\bf q},\omega+i\bar \nu)=\int \dfrac{d^3k}{4\pi^3}f_{{\bf k}}
\Big[\dfrac{\bar\nu^2({\bf k,k-q})}{\omega+i\bar\nu({\bf k,k-q})}
\Xi(\omega+i\bar\nu({\bf k,k-q}))-$$$$-\dfrac{\bar\nu^2({\bf k+q,k})}
{\omega+i\bar\nu({\bf k+q,k})}
\Xi(\omega+i\bar\nu({\bf k+q,k}))\Big].
\eqno{(1.14)}
$$

In integrals (1.10) -- (1.14) following designations are accepted
$$
\bar \nu({\bf k,k-q})=\dfrac{\nu({\bf k})+\nu({\bf k-q})}{2},
$$
$$
\bar \nu({\bf k+q,k})=\dfrac{\nu({\bf k+q})+\nu({\bf k})}{2},
$$
$$
\Xi(\omega+i\bar\nu({\bf k,k-q})=\dfrac{1}{\E_{\bf k-q}-\E_{{\bf k}}+
\hbar[\omega+i\bar\nu({\bf k,k-q})]},
$$
$$
\Xi(\omega+i\bar\nu({\bf k+q,k})=\dfrac{1}{\E_{\bf k}-\E_{{\bf k+q}}+
\hbar[\omega+i\bar\nu({\bf k+q,k})]}.
$$

\begin{center}
  \bf 2.  Longitudinal dielectric function of the quantum
  collisional degenerate plasmas with frequency of collisions,
  proportional to the module of a wave vector
\end{center}

Let's consider the frequency of collisions proportional
to the momentum module, or, that all the same, to the module of a wave vector
$$
\nu({\bf k})=\nu_0|{\bf k}|.
$$

Then
$$
\bar\nu({\bf k_1,k_2})=\dfrac{\nu({\bf k}_1)+\nu({\bf k}_2)}{2}=
\dfrac{\nu_0}{2}\Big(|{\bf k}_1|+|{\bf k}_2|\Big)
$$
and
$$
\bar \nu({\bf k,q})=\bar \nu\Big({\bf k}+\dfrac{{\bf q}}{2},
{\bf k}-\dfrac{{\bf q}}{2} \Big)=\dfrac{\nu_0}{2}\Big(\Big|{\bf k}+
\dfrac{{\bf q}}{2}\Big|+\Big|{\bf k}-\dfrac{{\bf q}}{2}\Big|\Big).
$$

The quantity $ \nu_0$ we take in the form $\nu_0=\dfrac{\nu}{k_F}$, where
$k_F$ is the Fermi wave number, $k_F=\dfrac{mv_F}{\hbar}$,
$ \hbar $ is the Planck's constant, $v_F$ is the Fermi electron velocity. Now
we have
$$
\nu({\bf k})=\dfrac{\nu}{k_F}|{\bf k}|.
\eqno{(2.1)}
$$

Let's notice, that on Fermi's surface, i.e. at $k=k_F $:
$ \nu (k_F) = \nu $. So, further in formulas (1.1) -- (1.7) frequency
collisions according to (2.1) it is equal
$$
\bar\nu({\bf k,q})=\dfrac{\nu}{2k_F}\Big(\Big|{\bf k}+
\dfrac{{\bf q}}{2}\Big|+\Big|{\bf k}-\dfrac{{\bf q}}{2}\Big|\Big).
\eqno{(2.2)}
$$

Instead of a vector $ {\bf k} $ we will enter the new dimensionless wave
vector of integration
$$
{\bf K}=\dfrac{{\bf k}}{k_F},\qquad d^3k=k_F^3\,d^3K.
$$

Let's enter also  new wave vector
$$
{\bf Q}=\dfrac{{\bf q}}{k_F}.
$$

At the specified replacement of variables we have
$$
f_{{\bf k}}=\Theta(\E_{\rm F}-\E_{{\bf k}})=\Theta(\E_{\rm F}-
\dfrac{\hbar^2{\bf k}^2}{2m})=\Theta(\E_{\rm F}-\dfrac{\hbar^2k_F^2}{2m}
{\bf K}^2)=
$$
$$
=\Theta(\E_{\rm F}-\E_{\rm F}{\bf K}^2)=\Theta(1-{\bf K}^2)=f_{\rm {\bf K}}.
$$

Here $\Theta(x)$ is the Heaviside function,
$$
\Theta(x)=\Bigg\{\begin{array}{c}
                   1,\qquad x>0, \\
                   0,\qquad x<0.
                 \end{array}
$$

According to the specified replacement of variables further it is received
$$
\bar \nu({\bf k,k-q})=\dfrac{\nu}{2}\Big(|{\bf K}|+|{\bf K-Q}|\Big),
$$
$$
\bar \nu({\bf k+q,k})=\dfrac{\nu}{2}\Big(|{\bf K+Q}|+|{\bf K}|\Big),
$$

$$
\E_{{\bf k-q}}-\E_{{\bf k}}+\hbar[\omega+i\bar\nu({\bf k,k-q})]=
$$
$$
=\dfrac{\hbar^2}{2m}\Big[({\bf k-q})-{\bf k}^2\Big]+
\hbar[\omega+i\bar\nu({\bf k,k-q})]=
$$
$$
=-2\E_{\rm F}Q\Big(K_x-\dfrac{Q}{2}\Big)+\hbar[\omega+i\bar\nu({\bf k,k-q})]=
$$
$$
=-2\E_{\rm F}Q\Big(K_x-\dfrac{Q}{2}-\dfrac{z^-}{Q}\Big).
$$

Here
$$
{\bf Q}=Q(1,0,0),\qquad z^-=x+iy\rho^-,\qquad
x=\dfrac{\omega}{k_{\rm F}v_{\rm F}},\qquad y=\dfrac{\nu}{k_{\rm F}v_{\rm F}},
$$

$$
\rho^-=\dfrac{1}{2}\Big(|{\bf K}|+|{\bf K-Q}|\Big)=
$$
$$
=\dfrac{1}{2}\Big[\sqrt{K_x^2+K_y^2+K_z^2}+
\sqrt{(K_x-Q)^2+K_y^2+K_z^2}\,\Big].
$$

Similarly we receive, that
$$
\E_{{\bf k}}-\E_{{\bf k+q}}+\hbar[\omega+i\bar\nu({\bf k,k-q})]=
$$

$$
=-2\E_{\rm F}Q\Big(K_x+\dfrac{Q}{2}-\dfrac{z^+}{Q}\Big),\qquad
z^+=x+iy\rho^+,
$$

$$
\rho^+=\dfrac{1}{2}\Big(|{\bf K}|+|{\bf K+Q}|\Big)=
$$
$$
=\dfrac{1}{2}\Big[\sqrt{K_x^2+K_y^2+K_z^2}+
\sqrt{(K_x+Q)^2+K_y^2+K_z^2}\,\Big].
$$

Let's pass to new variables in integrals (1.10) -- (1.14).
We receive following equalities. For integral (1.10) it is had
$$
B({\bf q},\omega+i\bar\nu)=-\dfrac{k_F^3}{8\pi^3\E_FQ}B(Q,z^{\pm}),
$$
where
$$
B(Q,z^{\pm})=\int f_{{\bf K}}\Big[\dfrac{1}{K_x-Q/2-z^-/Q}-\dfrac{1}
{K_x+Q/2-z^+/Q}\Big]d^3K.
$$

For integral (1.11) it is received
$$
b({\bf q},\omega+i\bar\nu)=-\dfrac{yk_F^3}{8\pi^3\E_FQ}b(Q,z^{\pm}),
$$
where
$$
b(Q,z^{\pm})=\int f_{{\bf K}}\Big[\dfrac{\rho^-}{z^-(K_x-Q/2-z^-/Q)}-
\dfrac{\rho^+}{z^+(K_x+Q/2-z^+/Q)}\Big]d^3K.
$$

For integral (1.12) it is received
$$
b({\bf q},0)=-\dfrac{yk_F^3}{8\pi^3\E_FQ}b(Q,0^{\pm}),
$$
where
$$
b(Q,0^{\pm})=\int f_{{\bf K}}\Big[\dfrac{\rho^-}{z^-(K_x-Q/2)}-
\dfrac{\rho^+}{z^+(K_x+Q/2)}\Big]d^3K.
$$

For integral (1.13) it is received
$$
b_{\bar\nu}({\bf q},\omega+i\bar\nu)=-\dfrac{yk_F^4v_F}{8\pi^3\E_{\rm F}Q}
b_{\bar\nu}(Q,z^{\pm}),
$$
where
$$
b_{\bar\nu}(Q,z^{\pm})=\int f_{{\bf K}}\Big[\dfrac{\rho^-}{K_x-Q/2-z^-/Q}-
\dfrac{\rho^+}{K_x+Q/2-z^+/Q}\Big]d^3K.
$$

At last, for integral (1.14) it is similarly received
$$
b_{\omega,\bar\nu}({\bf q},\omega+i\bar\nu)=-\dfrac{y^2k_F^4v_F}{8\pi^3\E_FQ}
b_{\omega,\bar\nu}(Q,z^{\pm}),
$$
where
$$
b(Q,z^{\pm})=\int f_{{\bf K}}\Big[\dfrac{{\rho^-}^2}{z^-(K_x-Q/2-z^-/Q)}-
\dfrac{{\rho^+}^2}{z^+(K_x+Q/2-z^+/Q)}\Big]d^3K.
$$

Let's substitute the received equalities in the formula (1.1). We receive
the expression for longitudinal dielectric function
$$
\varepsilon_l(Q,x,y)=1-\dfrac{3x_p^2}{4\pi Q^3}\Big[B(Q,z^{\pm})+
iyb_{\bar\nu}(Q,z^{\pm})\dfrac{b(Q,0)-b(Q,z^{\pm})}{xb(Q,0)+
iyb_{\omega,\bar\nu}(Q,z^{\pm})}\Big].
\eqno{(2.3)}
$$

Here $x_p$ is the dimensionless plasma (Langmuir) frequency,
$$
x_p=\dfrac{\omega_p}{k_{\rm  F}v_{\rm F}}, \qquad
\omega_p^2=\dfrac{4\pi^2 eN}{m},
$$
$\omega_p$ is the dimension plasma (Langmuir) frequency.

Let's notice, that in case of constant frequency of electron collisions
the quantity $ \rho^{\pm} $ passes in unit. Then we have
$$
B(Q,z^{\pm})=QB(Q,z), \qquad b(Q,0)=\dfrac{Q}{z}B(Q,0),
$$
$$
b_{\bar\nu}(Q,z^{\pm})=Q B(Q,z),\qquad b_{\omega,\bar\nu}(Q,z^{\pm})=
\dfrac{Q}{z}B(Q,z),
$$
where
$$
B(Q,z)=\int \dfrac{f_{{\bf K}}d^3K}{(K_x-z/Q)^2-(Q/2)^2}.
$$

Substituting these equalities in (2.3), we receive expression
of dielectric function for quantum degenerate
collisional plasmas with constant frequ\-ency of collisions
$$
\varepsilon_l(Q,x,y)=1-\dfrac{3x_p^2}{4\pi Q^2}B(Q,z)\Big[1+iy
\dfrac{B(Q,0)-B(Q,z)}{xB(Q,0)+iyB(Q,z)}\Big].
$$

Let's result the formula (2.3) in the calculation form. For this purpose
in the plane $ (K_y, K_z) $ we will pass to polar coordinates
$$
K_y^2+K_z^2=r^2,\qquad dK_ydK_z=rdrd\varphi.
$$

Then
$$
\varepsilon_l(Q,x,y)=1-$$$$-\dfrac{3x_p^2}{2Q^3}\Big[D(Q,z^{\pm})+
iyd_{\bar\nu}(Q,z^{\pm})\dfrac{d(Q,0)-d(Q,z^{\pm})}{xd(Q,0)+
iyd_{\omega,\bar\nu}(Q,z^{\pm})}\Big].
\eqno{(2.4)}
$$

Here
$$
D(Q,z^{\pm})=\int\limits_{-1}^{1}dK_x\int\limits_{0}^{\sqrt{1-K_x^2}}
\Big(\dfrac{1}{K_x-Q/2-z^-/Q}-\dfrac{1}{K_x+Q/2-z^+/Q}\Big)rdr,
$$
$$
\rho^-=\dfrac{1}{2}\Big(\sqrt{(K_x-Q)^2+r^2}+\sqrt{K_x^2+r^2}\Big),
$$
$$
\rho^+=\dfrac{1}{2}\Big(\sqrt{(K_x+Q)^2+r^2}+\sqrt{K_x^2+r^2}\Big).
$$

Besides,
$$
d(Q,z^{\pm})=$$$$=\int\limits_{-1}^{1}dK_x\int\limits_{0}^{\sqrt{1-K_x^2}}
\Big(\dfrac{\rho^-}{z^-(K_x-Q/2-z^-/Q)}-
\dfrac{\rho^+}{z^+(K_x+Q/2-z^+/Q)}\Big)rdr,
$$
$$
d(Q,0)=
$$
$$
=\int\limits_{-1}^{1}dK_x\int\limits_{0}^{\sqrt{1-K_x^2}}
\Big[\dfrac{\rho^-}{(x+iy\rho^-)(K_x-Q/2)}-
\dfrac{\rho^+}{(x+iy\rho^+)(K_x+Q/2)}\Big]rdr,
$$
$$
d_{\bar\nu}(Q,z^{\pm})=\int\limits_{-1}^{1}dK_x\int\limits_{0}^{\sqrt{1-K_x^2}}
\Big(\dfrac{\rho^-}{K_x-Q/2-z^-/Q}-
\dfrac{\rho^+}{K_x+Q/2-z^+/Q}\Big)rdr,
$$
and, at last,
$$
d_{\omega,\bar\nu}(Q,z^{\pm})=
$$
$$
=\int\limits_{-1}^{1}dK_x\int\limits_{0}^{\sqrt{1-K_x^2}}
\Big(\dfrac{{\rho^-}^2}{z^-(K_x-Q/2-z^-/Q)}-
\dfrac{{\rho^+}^2}{z^+(K_x+Q/2-z^+/Q)}\Big)rdr.
$$

On Figs. 1-8 comparison of the real and imaginary parts
of dielectric function depending on quantity of the dimensionless
wave vector $Q $ (Figs. 1-4) and depending on the dimensionless
quantities of frequency of an electro\-mag\-ne\-tic field $x $
(Figs. 5-8) is shown.
Thus curves 1 and 2 correspond to values of dimensionless frequency
collisions $y=0.1$ and $y=0.01$. Everywhere more low $x_p=1$.

On Figs. 9 and 10 comparison of  relative deviation
of real (curves 1) and imaginary parts (curves 2)
of dielectric function from the present work (with frequency of collisions,
proportional to the module of a wave vector) with the corresponding
parametres of dielectric Mermin function (with constant frequency
collisions) at the same parametres, and quantity
$y =\dfrac{\nu}{k_Fv_F} =0.01$ is the same. The last means, that on border
Fermi's surfaces quantity of frequency of collisions in both
dielectric functions is the same. Curves 1 on Figs. 9 and 10
are defined by function
$$
O_r(Q,x,y)=\dfrac{\Re \varepsilon_l^{\rm Mermin}(Q,x,y)-\Re\varepsilon_l(Q,x,y)}
{\Re \varepsilon_l^{\rm Mermin}(Q,x,y)},
$$
and curves  2 defined by function
$$
O_i(Q,x,y)=\dfrac{\Im \varepsilon_l^{\rm Mermin}(Q,x,y)-\Im\varepsilon_l(Q,x,y)}
{\Im \varepsilon_l^{\rm Mermin}(Q,x,y)}.
$$

On Figs. 11-14 comparison of the real and imaginary parts of
dielectric function according to frequency of collisions proportional
to the module of a wave vector (curves 1) and constant frequency
of collisions (curves 2) is shown.

\begin{center}
\bf 5. Conclusions
\end{center}

In the present work the formula for longitudinal
dielectric permeability into quantum collisional
degenerate plasma  is deduced. Comparison of the real and imaginary parts of
dielectric function  at various parametres is shown.

\clearpage

\begin{figure}[ht]\center
\includegraphics[width=16.0cm, height=10cm]{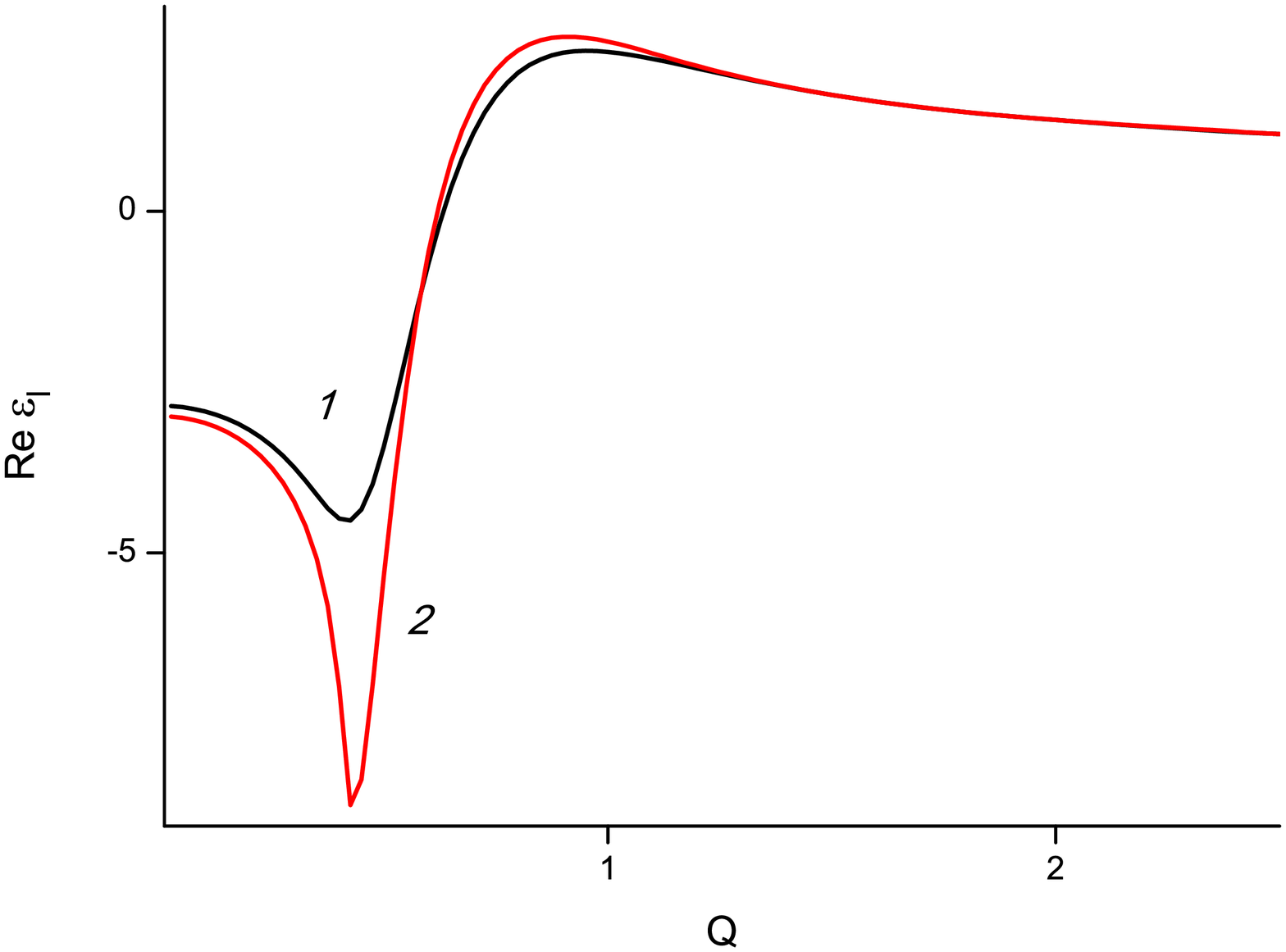}
\center{Fig. 1. Real part of dielectric function, $x=0.5$.
Curves $1,2$ correspond to values of dimensionless collision frequency
$y=0.1, 0.01$.}
\includegraphics[width=17.0cm, height=10cm]{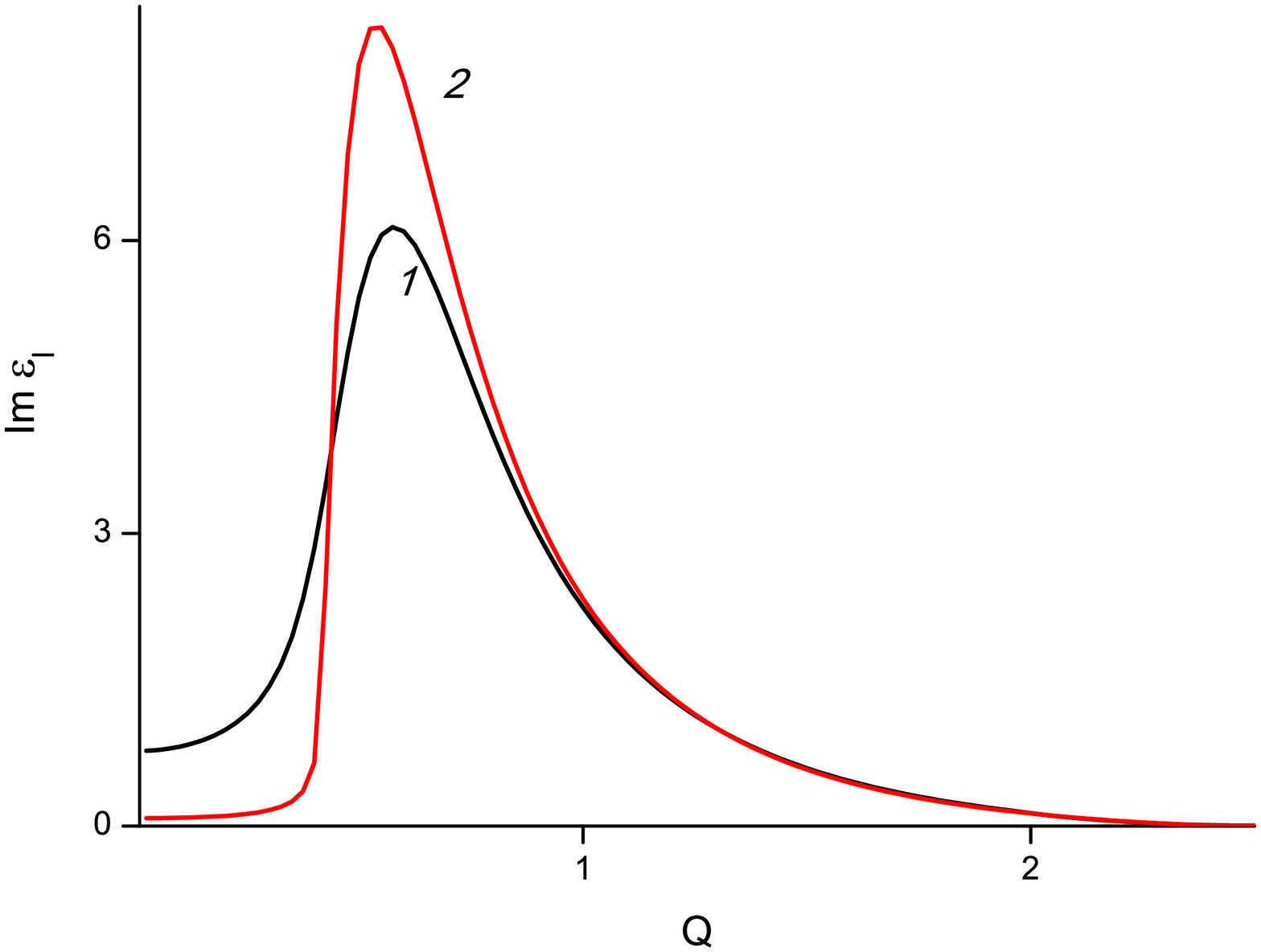}
\center{Fig. 2. Imaginary part of dielectric function, $x=0.5$.
Curves $1,2$ correspond to values of dimensionless collision frequency
$y=0.1, 0.01$.}
\end{figure}

\begin{figure}[ht]\center
\includegraphics[width=16.0cm, height=10cm]{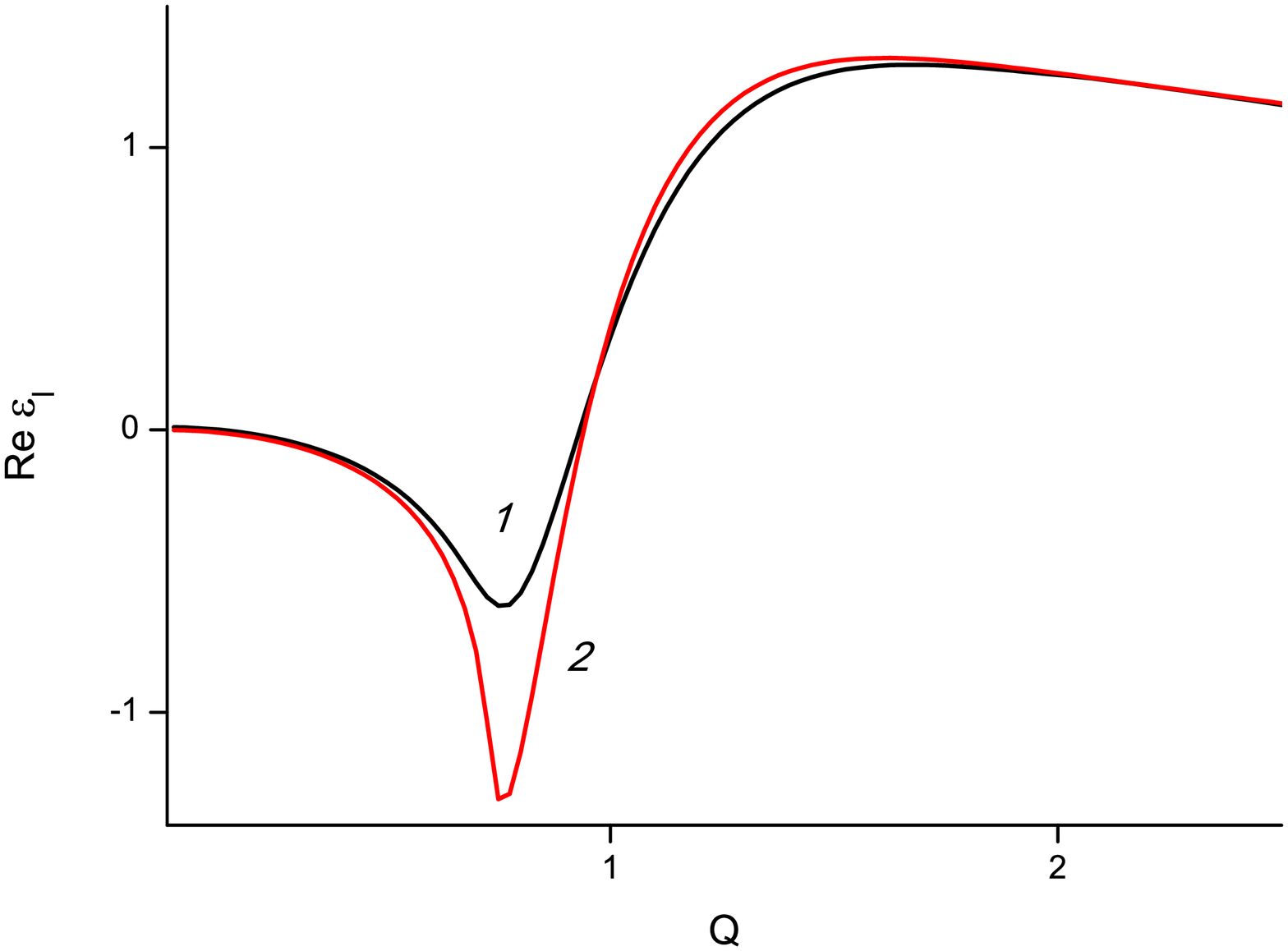}
\center{Fig. 3. Real part of dielectric function, $x=1$.
Curves $1,2$ correspond to values of dimensionless collision frequency
$y=0.1, 0.01$.}
\includegraphics[width=17.0cm, height=10cm]{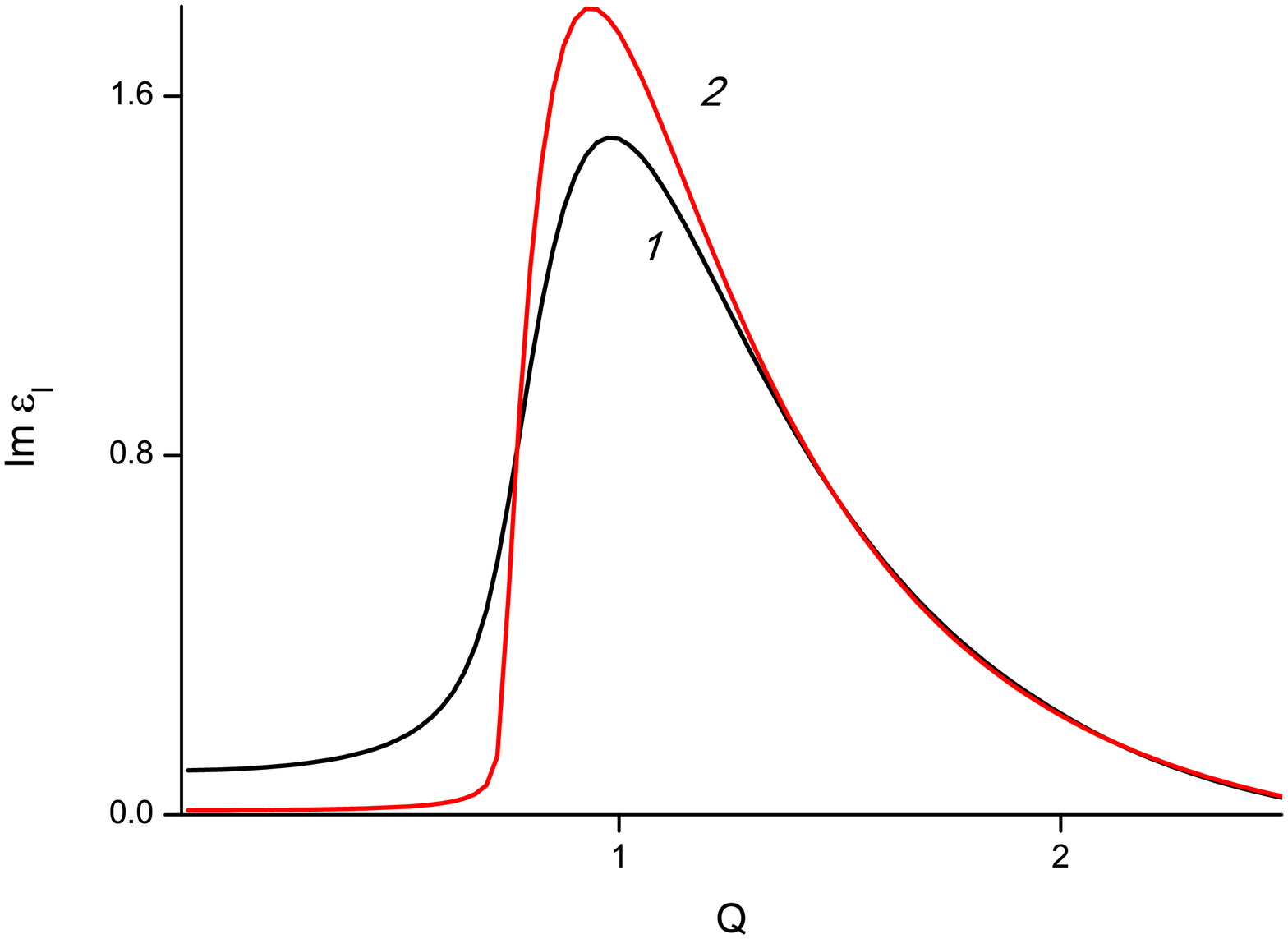}
\center{Fig. 4. Imaginary part of dielectric function, $x=1$.
Curves $1,2$ correspond to values of dimensionless collision frequency
$y=0.1, 0.01$.}
\end{figure}

\begin{figure}[ht]\center
\includegraphics[width=16.0cm, height=10cm]{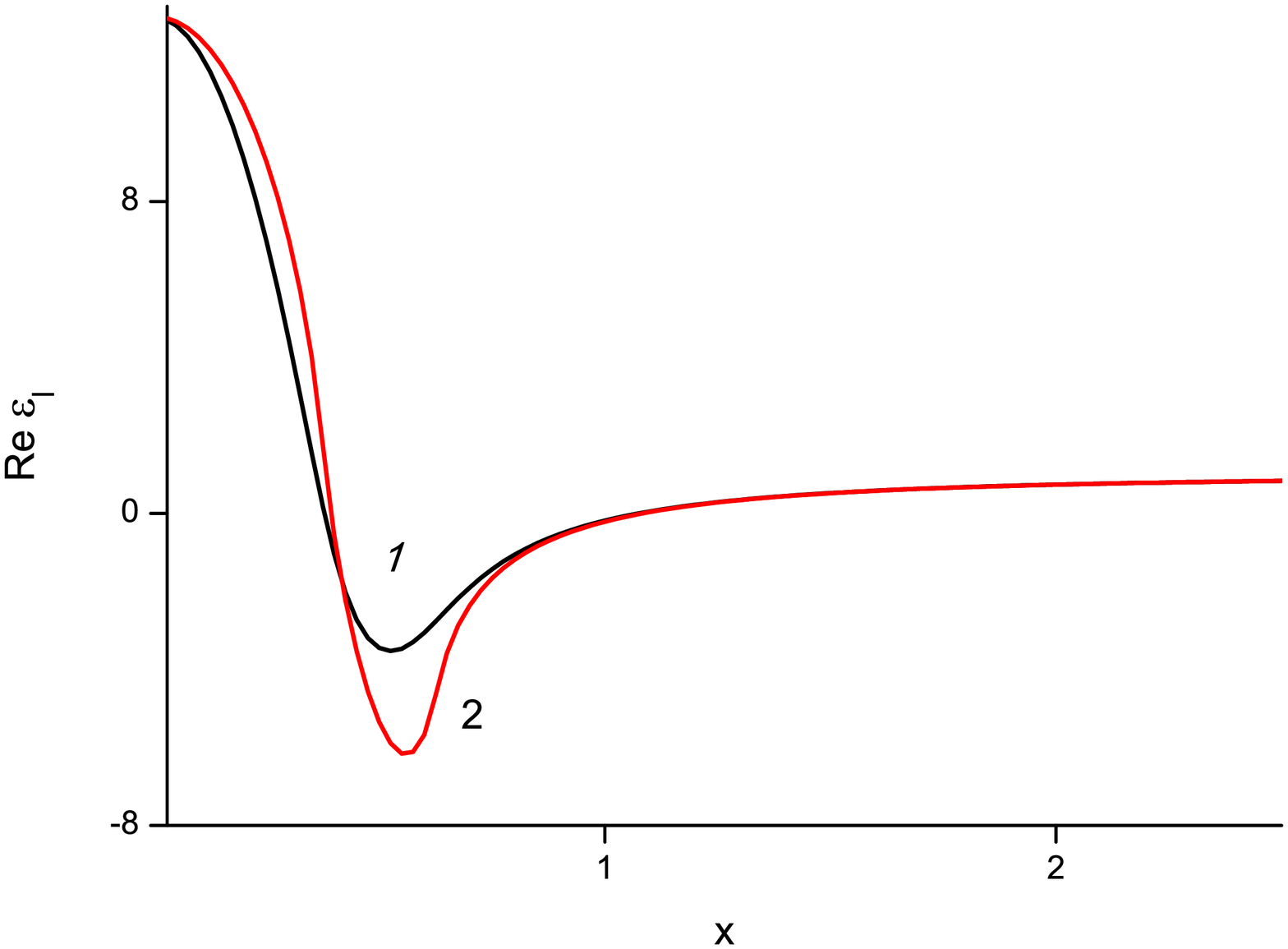}
\center{Fig. 5. Real part of dielectric function,  $Q=0.5$.
Curves $1,2$ correspond to values of dimensionless collision frequency
$y=0.1, 0.01$.}
\includegraphics[width=17.0cm, height=10cm]{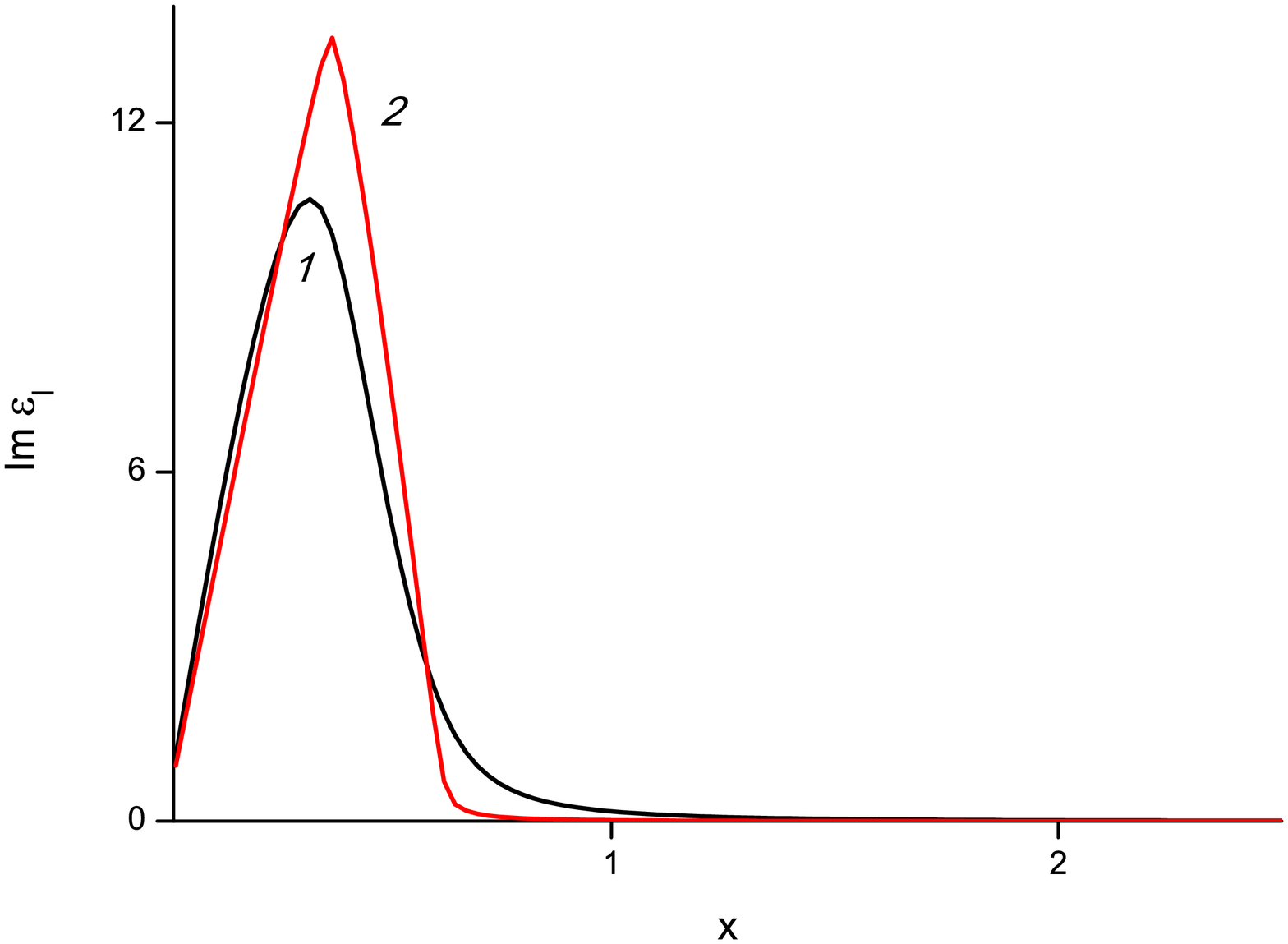}
\center{Fig. 6. Imaginary part of dielectric function, $Q=0.5$.
Curves $1,2$ correspond to values of dimensionless collision frequency
$y=0.1, 0.01$.}
\end{figure}

\begin{figure}[ht]\center
\includegraphics[width=16.0cm, height=10cm]{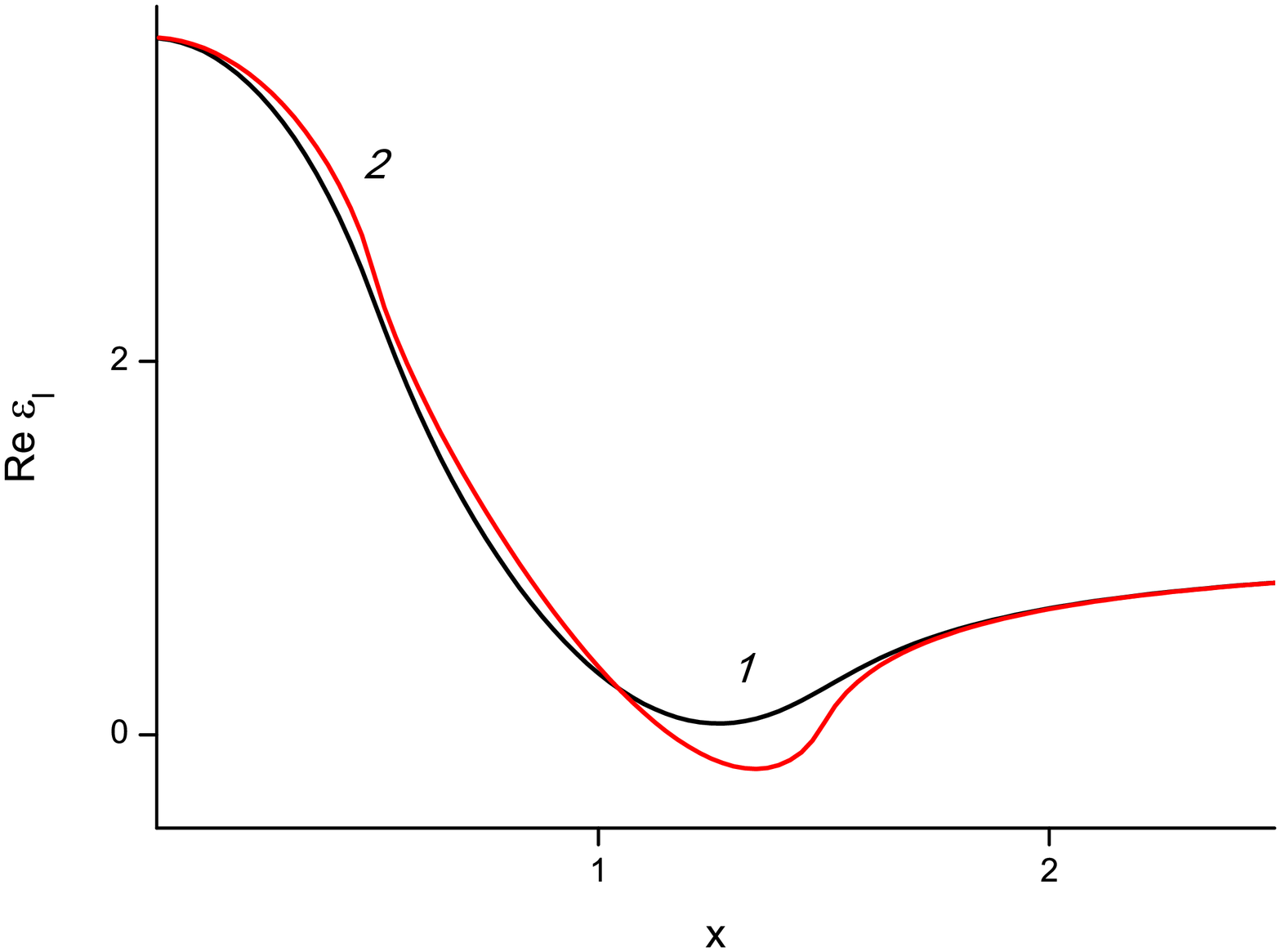}
\center{Fig. 7. Real part of dielectric function, $Q=1$.
Curves $1,2$ correspond to values of dimensionless collision frequency
$y=0.1, 0.01$.}
\includegraphics[width=17.0cm, height=10cm]{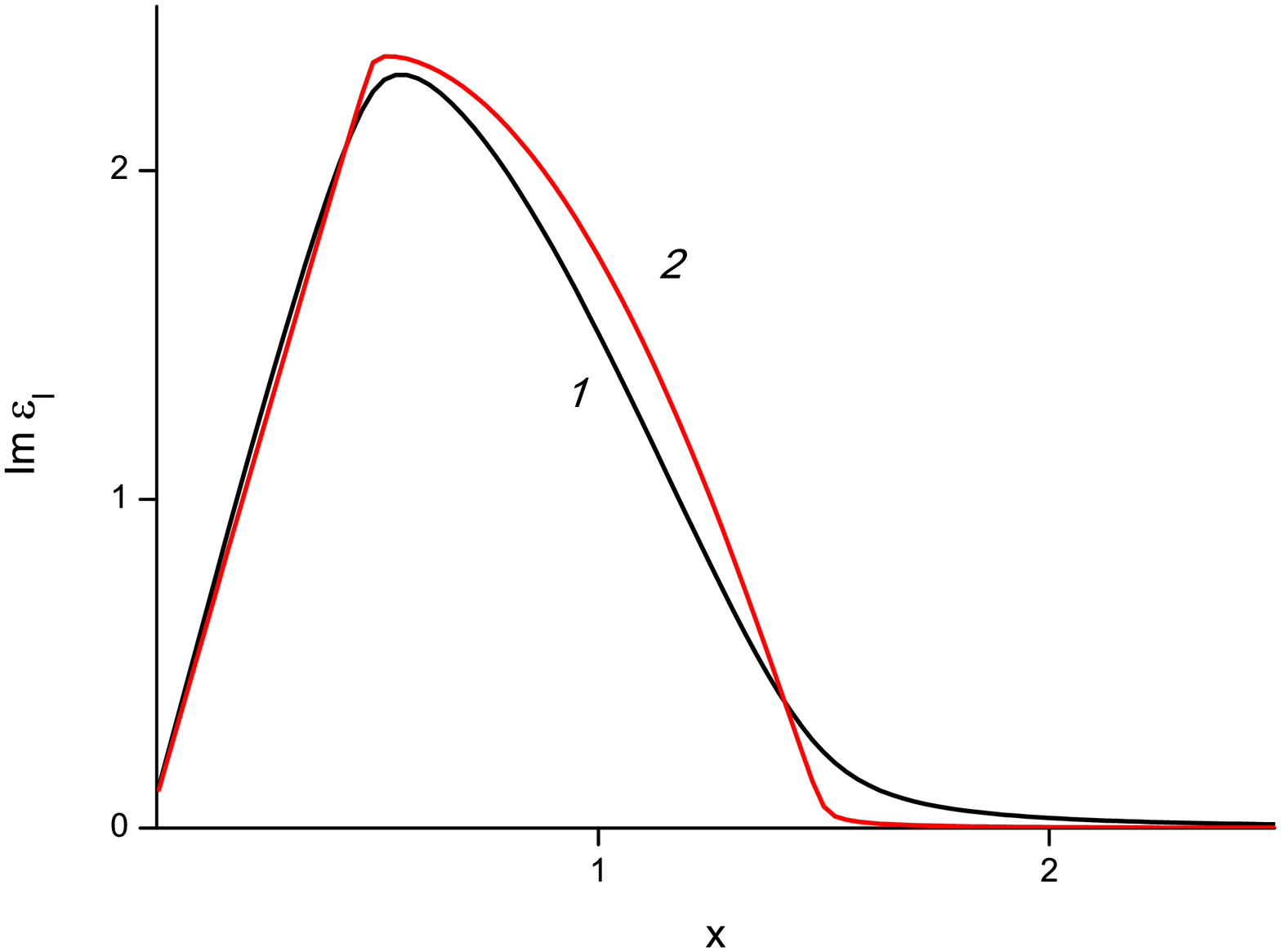}
\center{Fig. 8. Imaginary part of dielectric function, $Q=1$.
Curves $1,2$ correspond to values of dimensionless collision frequency
$y=0.1, 0.01$.}
\end{figure}

\begin{figure}[ht]\center
\includegraphics[width=16.0cm, height=10cm]{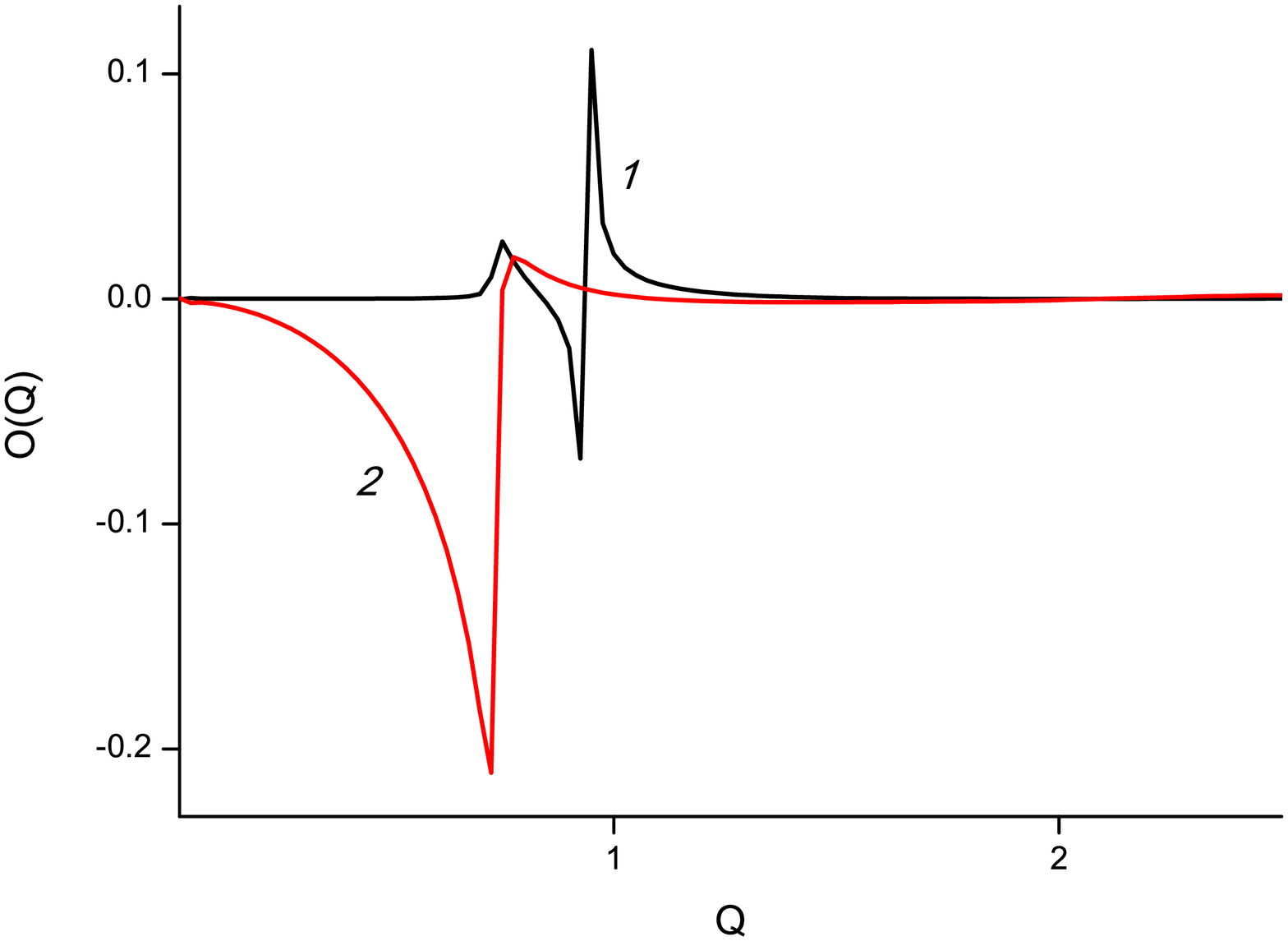}
\center{Fig. 9. Relative deviation of the real part
of dielectric function, $x=1$.
Curves $1,2$ correspond to values of dimensionless collision frequency
$y=0.1, 0.01$.}
\includegraphics[width=17.0cm, height=10cm]{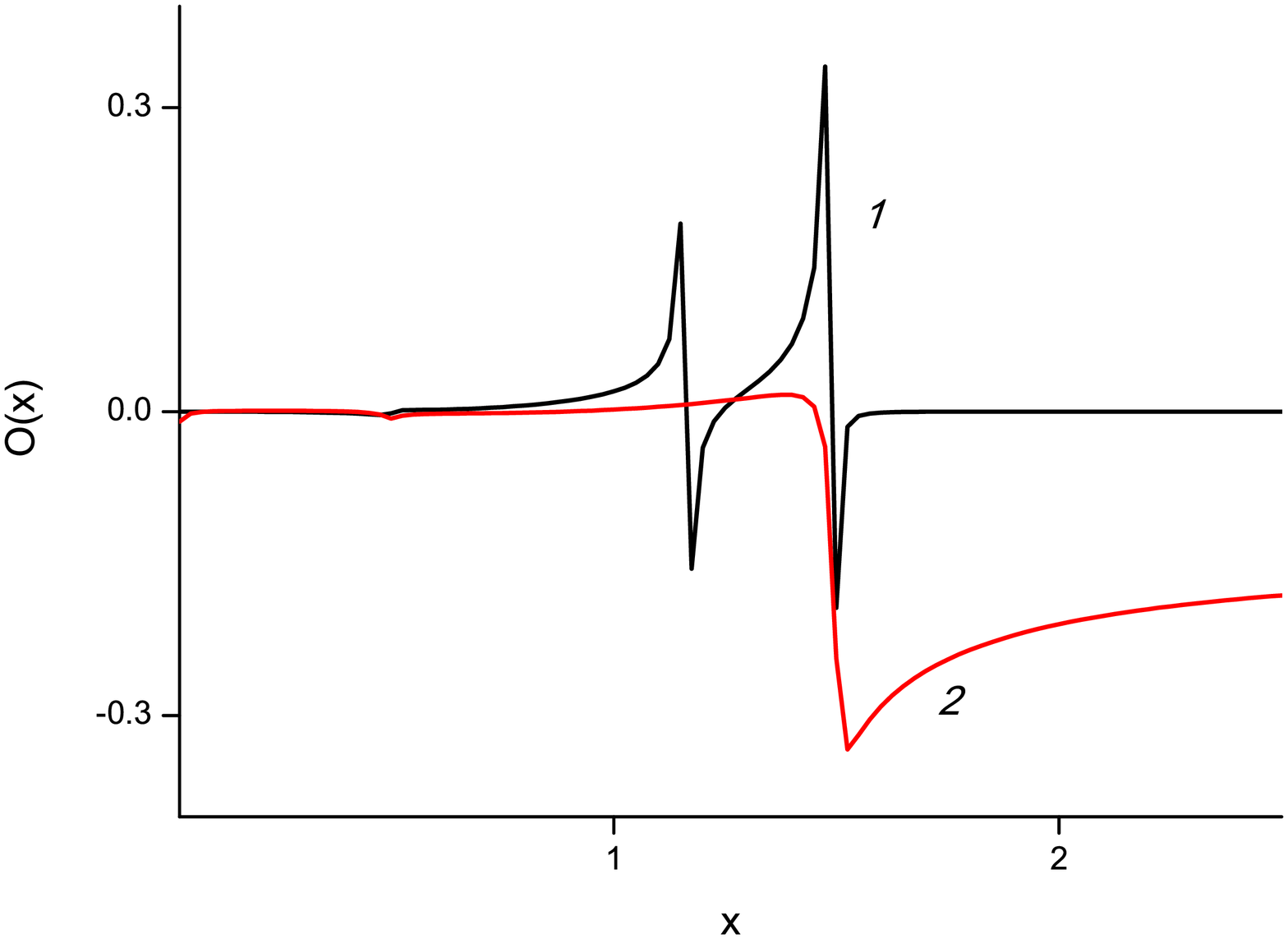}
\center{Fig. 10. Relative deviation of the imaginary part
of dielectric function, $Q=1$.
Curves $1,2$ correspond to values of dimensionless collision frequency
$y=0.1, 0.01$.}
\end{figure}

\begin{figure}[ht]\center
\includegraphics[width=16.0cm, height=10cm]{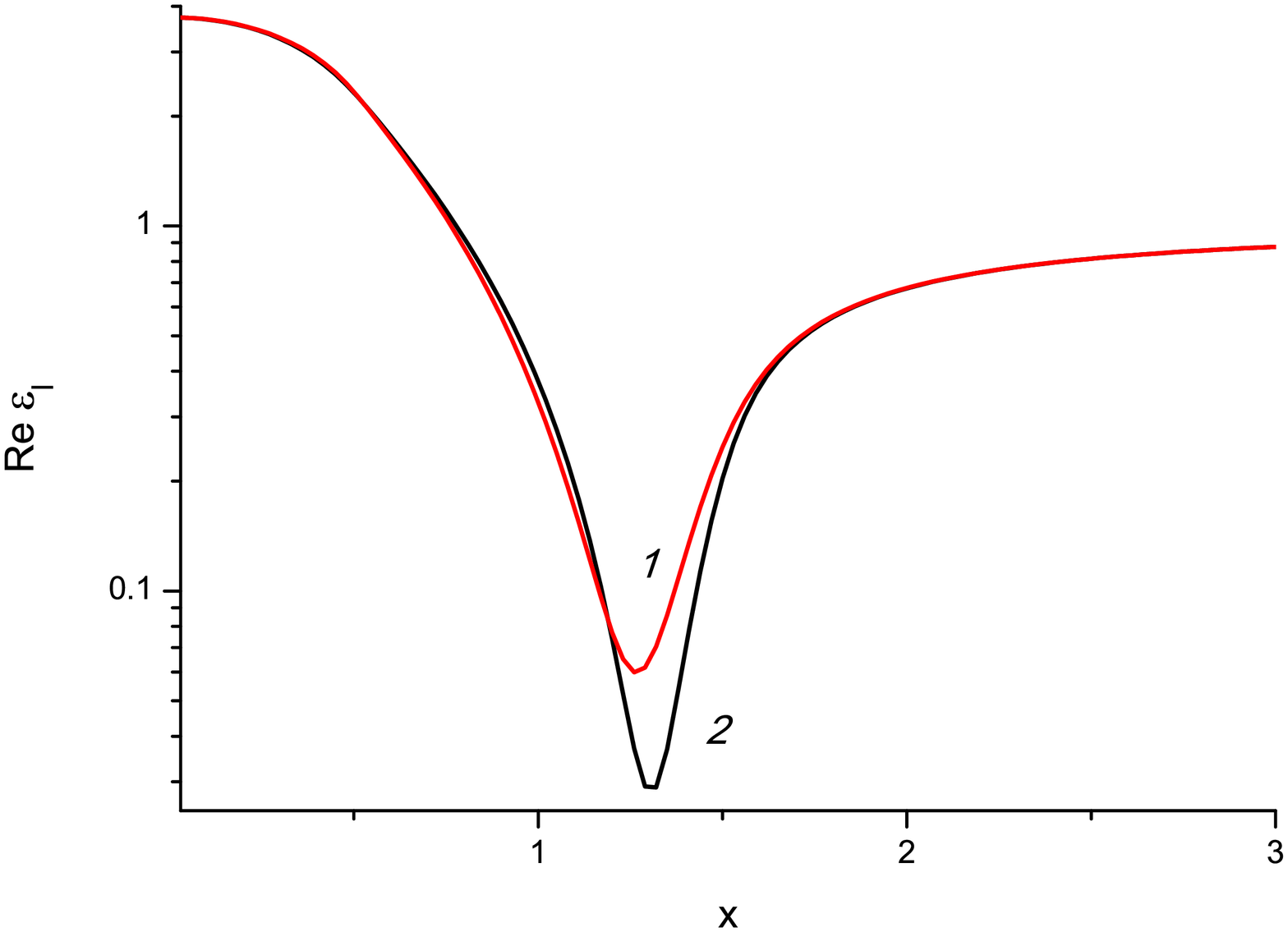}
\center{Fig. 11. Real part of dielectric function,
$Q=1, y=0.1$. Curves 1 and 2 correspond  accordingly  to variable and
constant collision frequency.}
\includegraphics[width=17.0cm, height=10cm]{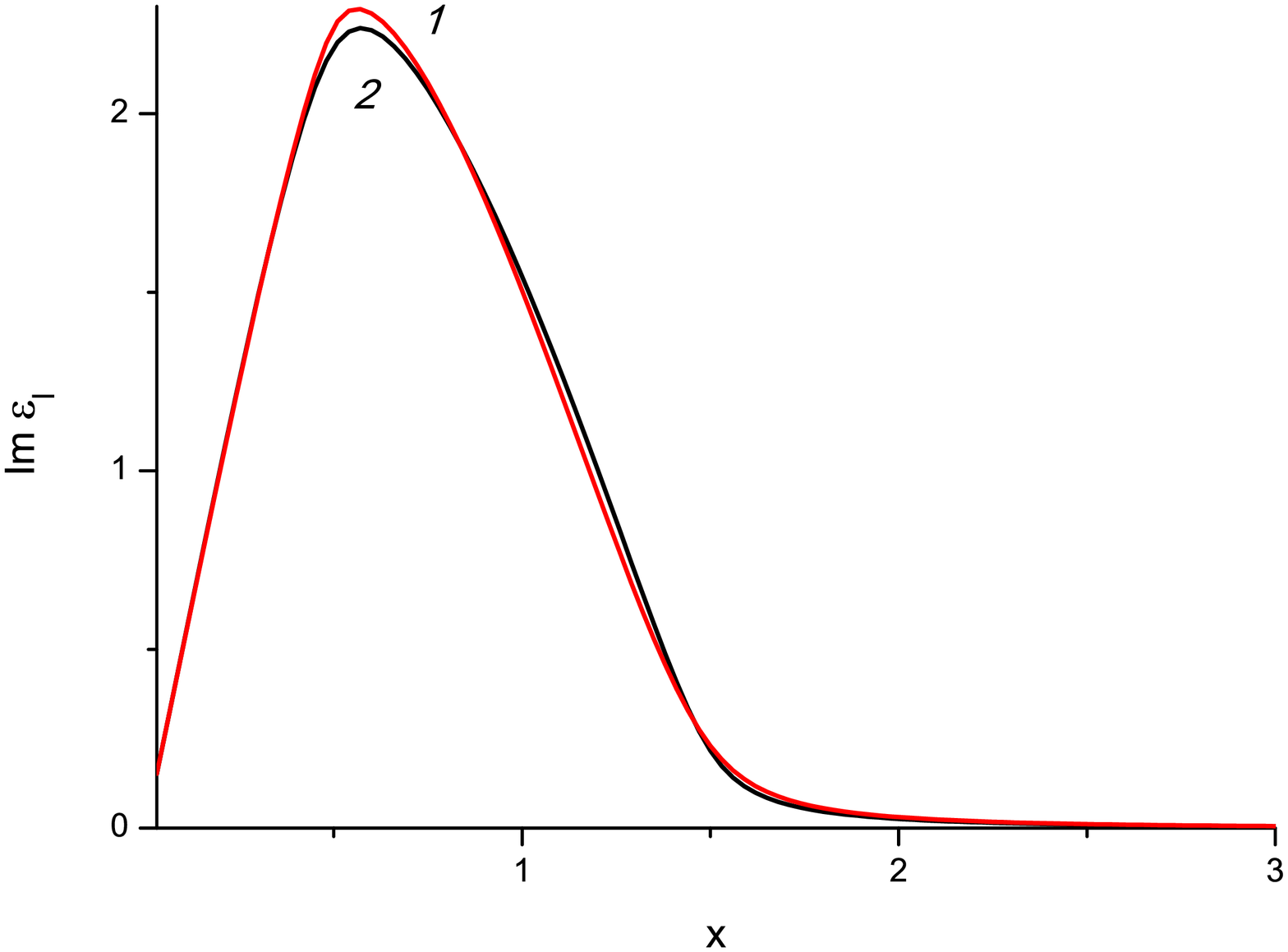}
\center{Fig. 12. Imaginary part of dielectric function,
$Q=1, y=0.1$. Curves 1 and 2 correspond  accordingly  to variable and
constant collision frequency.}
\end{figure}

\begin{figure}[ht]\center
\includegraphics[width=16.0cm, height=10cm]{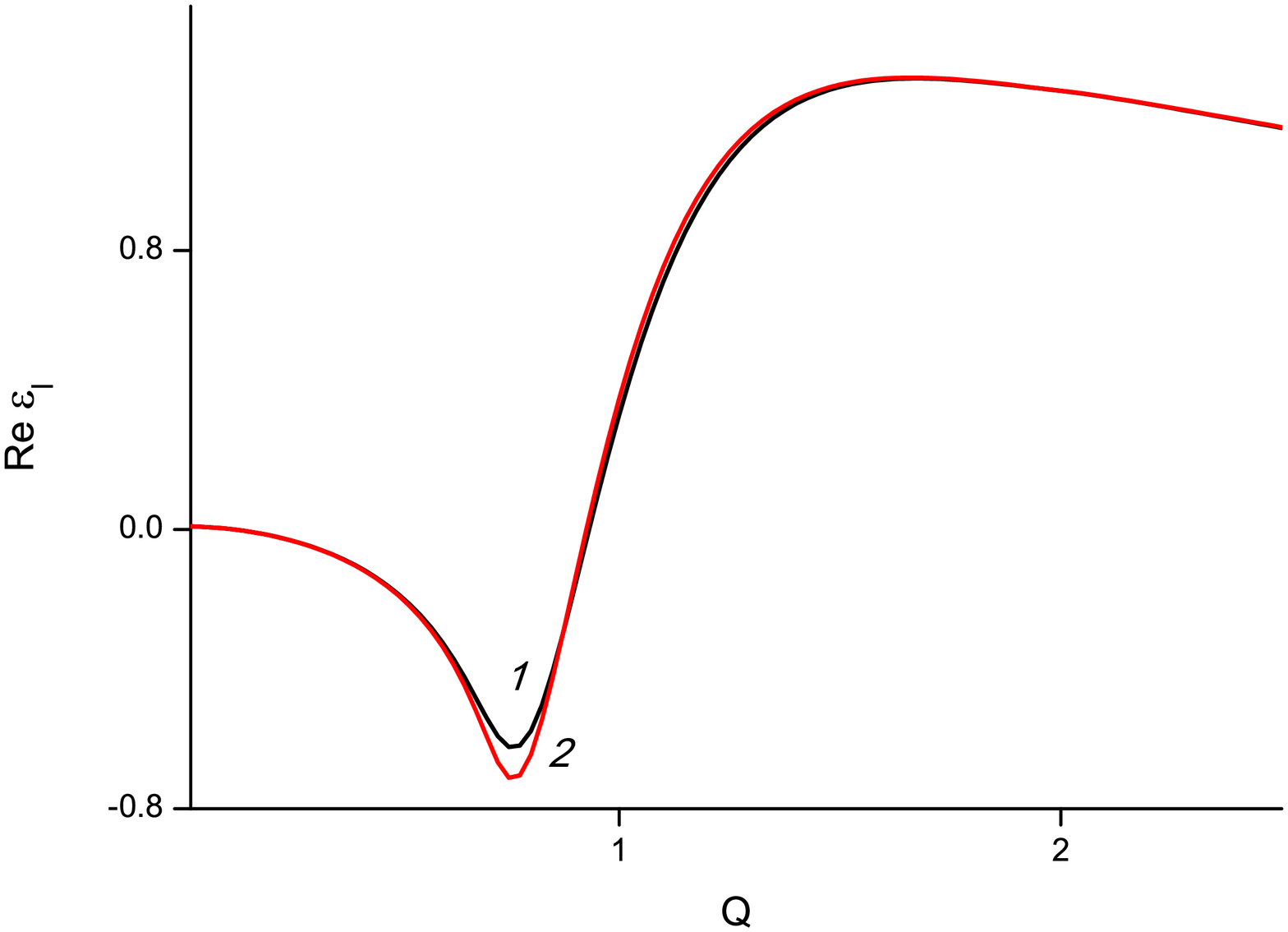}
\center{Fig. 13. Real part of dielectric function,
$x=1, y=0.1$. Curves 1 and 2 correspond  accordingly  to variable and
constant collision frequency.}
\includegraphics[width=17.0cm, height=10cm]{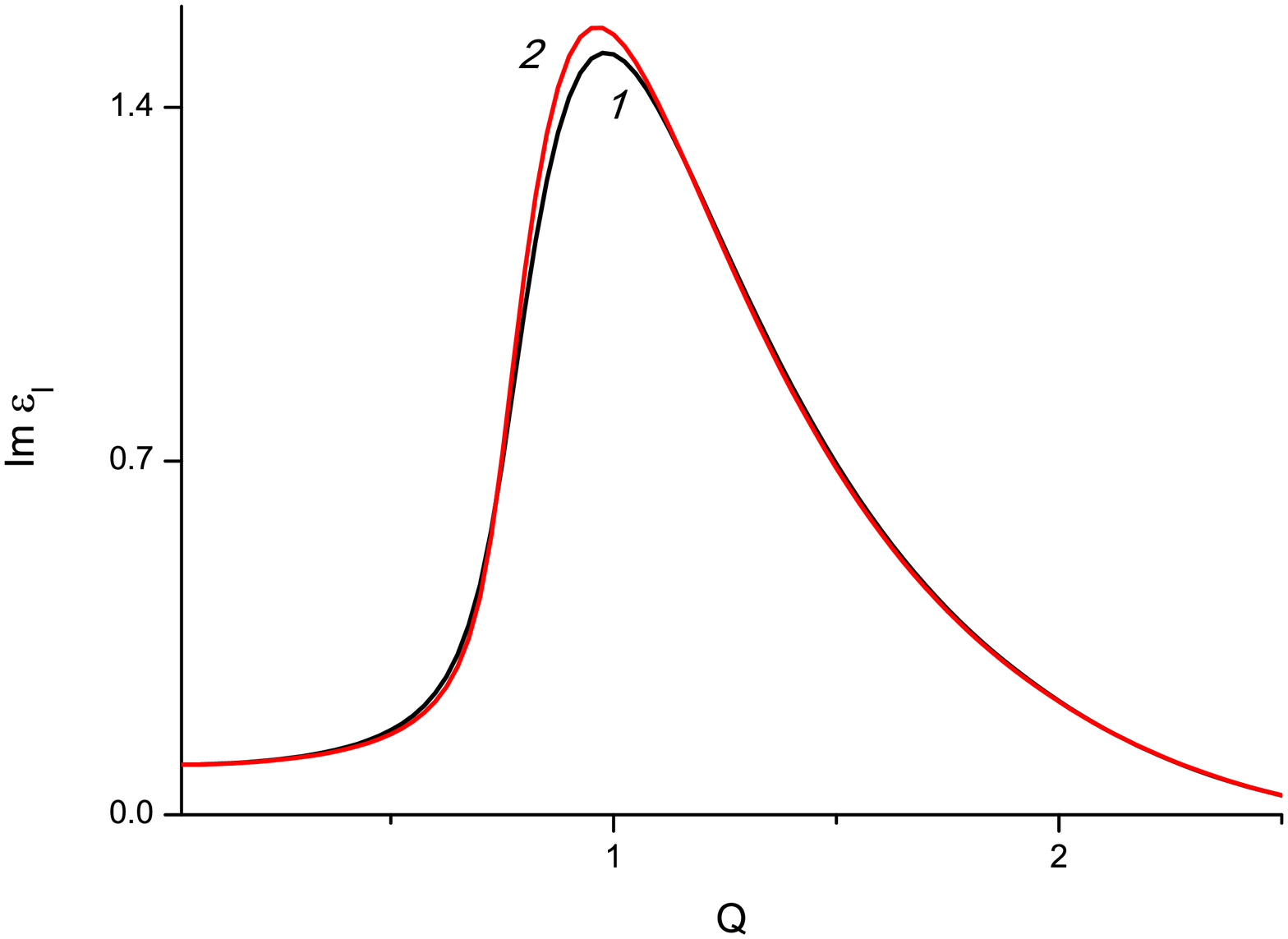}
\center{Fig. 14. Imaginary part of dielectric function,
$x=1, y=0.1$. Curves 1 and 2 correspond accordingly to  variable and
constant collision frequency.}
\end{figure}


\clearpage

\begin{thebibliography}{99}\normalsize
\renewcommand{\baselinestretch}{0.8}

\bibitem{Klim}{\it Klimontovich Y. and Silin V. P.} The Spectra of
Systems of Interacting Particles//
JETF (Journal Experimental Theoreticheskoi Fiziki), {\bf 23}, 151 (1952).


\bibitem{Lin} {\it Lindhard J.} On the properties of a gas of
charged particles// Kongelige Danske Videnskabernes Selskab,
Matematisk--Fysiske Meddelelser. V. 28, No. 8 (1954), 1--57.

\bibitem{Kliewer}{\it Kliewer K. L., Fuchs R.}
Lindhard Dielectric Functions with a Finite Electron Lifetime//
Phys. Rev. 1969. V. 181. No. 2. P. 552--558.


\bibitem{Mermin} {\it Mermin N. D.} { Lindhard Dielectric Functions
in the Relaxation--Time Approximation}.
Phys. Rev. B. 1970. V. 1, No. 5. P. 2362--2363.

\bibitem{Long} {\it Latyshev A.V., Yushkanov A.A.}
Longitudinal permettivity of a quantum degenerate
collisional plasma// Teor. and Mathem. Physics, {\bf 169}(3): 1739--1749
(2011).

\bibitem{Trans} {\it Latyshev A.V., Yushkanov A.A.}
Transverse Electric Conductivity in  Collisional Quantum Plasma//
Plasma Physics Reports, 2012, Vol. 38, No. 11, pp. 899--908.

\bibitem{Trans2}{\it Latyshev A.V., Yushkanov A.A.}
Transverse electric conductivity in quantum collisional
plasma in Mermin approach// arXiv:1109.6554v1 [math-ph]
29 Sep 2011.

\bibitem{Long2}{\it Latyshev A.V., Yushkanov A.A.}
Longitudinal electric conductivity and dielectric
permeability in quantum plasma with variable frequency of
collisions in Mermin' approach// arXiv:1212.5659v1
[physics.plasma-ph] 17 Jan 2013, 28 p.

\bibitem{Lat2007}
{\it Latyshev A.V., Yushkanov A.A.}
Transverse and Longitudinal Permi\-ti\-vi\-ties of a Gaseous
Plasma with an Electron Collision Frequency Pro\-por\-tio\-nal to the
Electron Velocity. --  Plasma Physics Report, 2007, Vol. 33, No. 8,
pp. 696--702 (Fizika Plasmy, Vol. 33, No. 8, pp. 762--768, russian).

\bibitem{Lat2006}
{\it Latyshev A.V., Yushkanov A.A.}
Skin Effect in a Gaseous Plasma with a Collision
Frequency Proportional to the Electron Velocity. --  Plasma
Physics Report. 2006. Vol. 32. No. 11, pp. 943 -- 948
(Fizika Plasmy, Vol. 32, No. 11, pp. 1021--1026, russian).

\bibitem{Manf} {\it Manfredi G.} { How to model quantum plasmas}//
arXiv: quant - ph/0505004.

\bibitem{Anderson} {\it Anderson D., Hall B., Lisak M., and
Marklund M.}
{Statistical effects in the multistream model for quantum plasmas}//
Phys. Rev. E {\bf 65} (2002), 046417.

\bibitem{Andres}{\it Andr\'{e}s P.,de, Monreal R., and
Flores F.}
{Relaxation--time effects in the transverse dielectric function and the
electromagnetic properties of metallic surfaces and small
particles}// Phys. Rev. {\bf B}. 1986. Vol. 34, No. 10, 7365--7366.

\bibitem{Shukla1} {\it Shukla P. K. and Eliasson B.}
Nonlinear aspects of quantum plasma physics//
Uspekhy Fiz. Nauk, {\bf 53}(1) 2010;[V. 180. No. 1, 55-82 (2010) (in Russian)].

\bibitem{Shukla2} {\it Eliasson B. and Shukla P.K.}
Dispersion properties of electrostatic oscillations in quantum
plasmas//
arXiv:0911.4594v1 [physics.plasm-ph] 24 Nov 2009, 9 pp.

\bibitem{Opher}{\it Opher M., Morales G. J., Leboeuf J. N.}
Krook collisional models of the kinetic susceptibility of plasmas//
Phys. Rev. E. V.66, 016407, 2002.

\bibitem{Gelder} {\it Gelder van, A.P.} Quantum Corrections in the
Theory of the Anomalous Skin Effect//
Phys. Rev. 1969. Vol. 187. No. 3. P. 833--842.

\bibitem{Fuchs}{\it Fuchs R., Kliewer K. L.}
Surface plasmon in a semi--infinite free--electron gas//
Phys. Rev. B. 1971. V. 3. No. 7. P. 2270--2278.

\bibitem{Fuchs2}{\it Fuchs R., Kliewer K. L.}
Optical properties of an
electron gas: further studies of a nonlocal description//
Phys. Rev. 1969. V. 185. No. 3. P. 905--913.

\bibitem{Dressel}{\it Dressel M., Gr\"{u}ner G.}
{Electrodynamics of Solids. Optical Properties of
Electrons in Matter}. - Cambridge. Univ. Press. 2003. 487 p.

\bibitem{Wier} {\it Wierling A.} {Interpolation between local
field corrections and the Drude model by a generalized Mermin
approach}//
arXiv:0812.3835v1 [physics.plasm-ph] 19 Dec 2008.

\bibitem{Brod} {\it Brodin G., Marklund M., Manfredi G.}
{Quantum Plasma Effects in the Classical Regime}//
Phys. Rev. Letters. {\bf 100}, (2008). P. 175001-1--175001-4.

\bibitem{Manf2} {\it Manfredi G. and Haas F.}
{Self-consistent fluid model for a quantum electron gas}//
Phys. Rev. B {\bf 64} (2001), 075316.

\bibitem{Ropke} {\it Reinholz H., R\"{o}pke G.} Dielectric function beyond
the random-phase approximation: Kinetic theory versus linear response
theory// Phys. Rev., {\bf E 85}, 036401 (2012).

\end{thebibliography}
\end{document}